\documentclass[letterpaper]{jhep3}

\usepackage{graphicx}
\usepackage{graphics}
\usepackage{amssymb}
\usepackage{amsthm}

\usepackage[]{caption} 
\usepackage{amsmath} 
\def\beq{\begin{equation}}
\def\eeq{\end{equation}}
\def\bea{\begin{eqnarray}}
\def\eea{\end{eqnarray}}

\newcommand{\cA}{{\cal A}}
\newcommand{\cAb}{{\overline{\cal A}}}
\newcommand{\cF}{{\cal F}}
\newcommand{\cFb}{{\overline{\cal F}}}
\newcommand{\cD}{{\cal D}}
\newcommand{\cDb}{{\overline{\cal D}}}
\newcommand{\KD}{{K\"{a}hler-Dirac }}
\newcommand{\cQ}{{\cal Q}}
\newcommand{\cU}{{\cal U}}
\newcommand{\cN}{{\cal N}}
\newcommand{\cUb}{{\overline{\cal U}}} 
\newcommand{\Tr}{{\rm Tr\;}}

\newcommand{\bx}{{\bf x}}

\def\bec{\begin{center}}
\def\eec{\end{center}}

\title{Thermal phases of D1-branes on a circle from lattice super Yang-Mills}

\author{Simon Catterall \\
Department of Physics, Syracuse University, Syracuse, NY13244, USA\\
E-mail: \email{smc@phy.syr.edu}
}

\author{Anosh Joseph \\
Department of Physics, Syracuse University, Syracuse, NY13244, USA\\
E-mail: \email{ajoseph@phy.syr.edu}
}

\author{Toby Wiseman \\
Theoretical Physics Group, Blackett Laboratory, Imperial College, London SW7 2AZ, UK\\
E-mail: \email{t.wiseman@imperial.ac.uk}
}

\preprint{SU-4252-911}

\date{}

\abstract{We report on the results of numerical simulations of
$1+1$ dimensional $SU(N)$ Yang-Mills theory with maximal
supersymmetry at finite
temperature and compactified on a circle. For large $N$  this
system is thought to provide a dual description
of the decoupling limit 
of $N$ coincident D1-branes on a circle. It has been proposed that at large $N$ there is a phase transition at strong coupling related to the Gregory-Laflamme (GL) phase transition in the holographic gravity dual. In a  high temperature limit there was argued to be a deconfinement transition associated to the spatial Polyakov loop, and it has been proposed that this is the continuation of the strong coupling GL transition.
Investigating the theory on the lattice for $SU(3)$ and $SU(4)$ and studying the time and space Polyakov loops we find evidence supporting this. In particular at strong coupling we see the transition has the parametric dependence on coupling predicted by gravity. We estimate the GL phase transition temperature from the lattice data which, interestingly, is not yet known directly in the gravity dual.  Fine tuning in the lattice theory
is avoided by the use of a lattice action with exact supersymmetry.}
 
\begin{document}

%
\section{Introduction}
%

In recent years a series of numerical studies have been undertaken to explore
and test conjectured holographic dualities between supersymmetric gauge
theories and supergravity theories \cite{Itzhaki:1998dd}. So far these studies have been
confined to the case when the super Yang-Mills theory is one dimensional and
the dual gravitational theory describes the low energy dynamics of D0-branes
\cite{Hanada:2007ti,Catterall:2007fp,Anagnostopoulos:2007fw,Catterall:2008yz,Hanada:2008gy,Hanada:2008ez,Catterall:2009xn,Hanada:2009ne} or the ${\cal N}=4$ theory compactified on $S^3\times R$ 
\cite{Catterall:2010gf,Ishiki:2008te,Ishiki:2009sg}

In this paper we extend these calculations to
the case of $N$ coincident D1-branes wrapped on a spatial circle, which in the decoupling limit is described by 
a two dimensional maximally supersymmetric
Yang-Mills (SYM) theory on a circle \cite{Itzhaki:1998dd,Aharony:2004ig}. (See \cite{Azeyanagi:2009zf, Hanada:2009hq, Hanada:2010kt, Hanada:2010gs} for some recent developments on the lattice along this direction.) This two dimensional Yang-Mills system possesses a richer structure at large $N$ than
its one dimensional counterpart, since when one compactifies the spatial direction on a circle, one can construct a new dimensionless coupling that can be varied in addition to the temperature. Arguments from a high temperature limit and also from strong coupling, using a dual  supergravity description, indicate that the system
should possess an interesting phase structure in the 2d parameter
space spanned by the
temperature and this new coupling in the large $N$ limit \cite{Aharony:2004ig,Aharony:2005ew}. A large $N$  transition between confined and deconfined phases with respect to the spatial Polyakov line is expected which interpolates between the high temperature region and the strongly coupled region. In particular for the strongly coupled region the dual D1-brane system can be described by certain black holes in supergravity, with a compact spatial circle. Then arguments from 
the dual gravity model indicate a first order Gregory-Laflamme (GL) phase transition between the black hole solutions localized on the circle
and uniform black hole solutions which wrap the circle \cite{Susskind:1997dr,Li:1998jy,Martinec:1998ja,Kol:2002xz,Aharony:2004ig,Harmark:2004ws,Aharony:2005ew,Harmark:2005pq}. Translating back to the SYM, the dual gravity model predicts the parametric dependence of the transition temperature against dimensionless circle coupling -- a dependence which seemingly cannot be deduced by simple SYM considerations. Interestingly, since the relevant gravity solutions have not been constructed yet (analog solutions are known, but not in the correct dimension \cite{Kudoh:2003ki,Kudoh:2004hs,Headrick:2009pv}), the precise coefficient in this relation is not known, and determining it in SYM yields a prediction for the phase transition temperature that could be tested in the future when the gravity solutions are constructed - a classical but nonetheless rather non-trivial gravitational problem. 

The purpose of this paper is to use Monte Carlo simulation of a lattice formulation 
of the two dimensional SYM theory to investigate its phase structure, focussing on possible large $N$ transitions between spatially
confined and deconfined phases of the model as revealed by
behavior of the spatial 
Polyakov line. In the next section we review the theoretical background to the
conjectured 2d Yang-Mills/D1-brane duality when the theories are compactified on a circle and describe the 
expected phase structure
in certain limits.
 
The usual problems associated with
the study of supersymmetric lattice theories are avoided by use of
new formulations which possess exact supersymmetric invariances at
non-zero lattice spacing - the relevant lattice construction is described in
section~\ref{sec:latticeact}.
We then describe our numerical results, which appear to confirm the expected deconfinement phase transition. Furthermore, at strong coupling the 
position of the observed critical
line agrees with the parametric dependence on couplings predicted by the dual gravity analysis. In particular we estimate the coefficient in this relation and hence derive a prediction for the GL phase transition temperature for the dual black holes theory. 
 We believe that this is the first time holography has made a detailed prediction for properties of some currently unknown non-trivial solutions in classical gravity based on calculations in strongly coupled Yang-Mills.
\\

\noindent
{\bf Note added in version 3:} In the previous versions of this paper the definition of the coupling in the lattice code was misidentified with that in~\eqref{eq:SYM}, which is the one appropriate for both the gravity and $0+1$ bosonic QM predictions given, instead being normalized as $S = \frac{N}{\lambda} \int_{T^2} d\tau dx \mathrm{Tr} \left( \tfrac{1}{2} F_{\mu\nu}^2 + \ldots \right)$. Thus there was a factor of two error in the continuum $\lambda$ when comparison was made to gravity and $0+1$ bosonic QM predictions. In particular this led to a $\sqrt{2}$ error in the identification of $c_{crit}$. This is corrected in the current version, with the normalization defined in~\eqref{eq:SYM}.
\footnote{The authors are very grateful to Paul Romatschke for discussion and comparison with his bosonic 1+1 code resulting in identification of this error in normalization.}

%
\section{Theoretical background}
\label{sec:theory}
%

We are interested in studying large $N$ thermal two dimensional maximally supersymmetric (16 supercharge) $SU(N)$ Yang-Mills theory, in the 't Hooft limit, with coupling $\lambda = N g_{YM}^2$, with the spatial direction compactified. Continuing the theory to Euclidean time, this implies the Yang-Mills theory is defined on a rectangular 2-torus, with time cycle size $\beta$, and space cycle size $R$. The fermion boundary conditions distinguish the two cycles, being anti-periodic on the time cycle so that $\beta$ has the interpretation of inverse temperature, and periodic boundary conditions on the space cycle. The action may then be written as,
\begin{eqnarray}
\label{eq:SYM}
S = \frac{N}{\lambda} \int_{T^2} d\tau dx \mathrm{Tr} \left( \tfrac{1}{4} F_{\mu\nu}^2 + \tfrac{1}{2} \sum_I [ D_\mu \phi^I , D^\mu \phi^I ]^2 - \tfrac{1}{4} \sum_{I , J} [ \phi^I , \phi^J ]^2 + \mathrm{fermions} \right)
\end{eqnarray}
where $I,J = 1,\ldots,8$ and $\phi^I$ are the 8 adjoint scalars, and $\tau$ is the coordinate on the time circle, and $x$ the coordinate on the space circle. Since $\lambda$, $\beta$ and $R$ are dimensionful, it is convenient to work with the two dimensionless couplings, $r_\tau = \sqrt{\lambda} \, \beta$ and $r_x = \sqrt{\lambda} \, R$ which give the dimensionless radii of the time and space circles respectively, measured in units of the 't Hooft coupling. We will be interested in the expectation values of the trace of the Polyakov loops on the time and space circles,
\beq
P_{\tau} = \frac{1}{N} \Big{\langle} \left| \Tr (P \exp (i \oint A_{\tau})) \right| \Big{\rangle}~, \; P_{x} =  \frac{1}{N} \Big{\langle}  \left| \Tr (P \exp (i \oint A_{x}))  \right| \Big{\rangle}~,
\eeq
as at large $N$, these give order parameters for confinement/deconfinement (or center symmetry breaking) phase transitions which we will discuss below.

As discussed in \cite{Aharony:2004ig,Aharony:2005ew} there are several interesting limits for the theory. In the large torus limit, $1 \ll r_x, r_\tau$ the string theory dual may be described by supergravity. For the weak coupling limit, $r_x , r_\tau \ll 1$, or asymmetric torus limits $r_\tau \ll r_x^3$ and $r_x \ll r_\tau^3$, we will find the dynamics is captured by a lower dimensional YM theory. Let us now review these cases and their predictions.

\subsection{Large torus limits and IIB and IIA supergravity duals}

When the torus becomes large in units of the 't Hooft coupling one finds that in certain regimes the dual D1-branes in string theory can be well described by supergravities \cite{Itzhaki:1998dd} as we shall now briefly review. Having a supergravity description of the full string theory dual allows certain behaviours of the theory to be studied using simple semi-classical gravity reasoning which allows powerful predictions to be inferred for the dual SYM.

\FIGURE[h]{
 \centerline{ 
 \includegraphics[width=.6\textwidth]{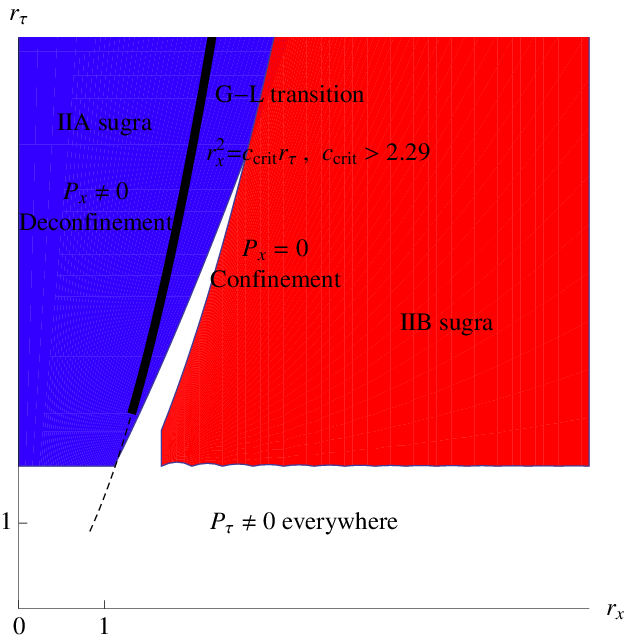}
 }
         \caption{Figure indicating the regions of coupling space where at large $N$ the dual string theory may be approximated by (red) IIB supergravity and (blue) IIA supergravity. In these regions, the SYM thermodynamics is dual to the thermodynamics of certain black holes in the corresponding supergravity. The IIA region predicts a large $N$ first order phase transition (the Gregory-Laflamme phase transition) between black holes localized on the spatial circle, and wrapping over the circle. The phase transition is known  to occur along the curve $r_x^2 = c_{crit} r_\tau$ where $c_{crit}$ is a constant, not yet determined, but known to be order one and $c_{crit} > 2.29$. The SYM transition is thought to be a deconfinement transition of the spatial Polyakov loop.}
   \label{fig:sugra}
}

The dual IIB string theory is given by the `decoupling limit' of $N$ coincident $D1$-branes \cite{Itzhaki:1998dd}. This decoupling limit is where one considers finite energy excitations of the $D1$-branes while taking the limit,~~$g^2_{YM} = \frac{1}{2\pi}\frac{g_s}{\alpha'} = \textrm{fixed},~~\alpha' \rightarrow 0$, where $g_s$ is the string coupling and $\alpha'$ determines the string tension. Since our Euclidean SYM is defined on a torus, the string dual is too, being at finite temperature and having one spatial direction compactified into a circle radius $R$ with periodic fermion boundary conditions.

One finds that for $1 \ll r_\tau \ll r_x^2$ this string theory can be described effectively by its supergravity sector. String oscillator and winding mode corrections to this supergravity description are small in this limit. The IIB supergravity solution describing the thermal vacuum is a black hole, carrying electric $D1$-brane charge. The $D1$-brane charge is string like (ie. its field strength tensor is a 3-form), and the appropriate configuration is to take the charge to wrap over the compact space circle. The solution preserves translational invariance around the space circle direction and is thought to be stable to small perturbations.

However there is a second supergravity description of the theory which is valid in a partly overlapping and partly complementary range $1 \ll r_\tau$ and $r_x^{4/3} \ll r_\tau$, obtained by performing a T-duality 
transformation on the compact spatial circle of the IIB string theory \cite{Aharony:2004ig,Aharony:2005ew}. Roughly speaking, such a T-duality exchanges winding and momentum modes of the string on this spatial circle, and exchanges the IIB string theory for a IIA string theory. In our case the $N$ $D1$-branes now get exchanged with $N$ $D0$-branes in the IIA theory. Since the $D0$-branes are point like, rather than string like, they have freedom to distribute their electric charge over the circle in various ways - it may be uniformly distributed, non-uniformly distributed, or fully localized on the circle, the latter two choices breaking the translational symmetry along the space circle direction. It is then a dynamical question which case is preferred.

It is thought \cite{Aharony:2004ig} that there are 3 types of black hole solution which indeed realize these 3 choices. The uniform black hole solution exists for all temperatures, but it is known to have a dynamical perturbative instability of the Gregory-Laflamme type \cite{Gregory:1993vy,Gregory:1994bj} for low temperatures $r_x^2 \leq 2.29 \, r_\tau$ \cite{Aharony:2004ig}. For higher temperatures it is thought to be dynamically stable. However, at a higher temperature than the instability point, so that
$r_x^2 = c_{crit} \, r_\tau$ for some constant $c_{crit}$ with $c_{crit} > 2.29$, the uniform black hole is thought to become globally thermodynamically less favored than the localized black hole solution. The actual transition temperature which governs the constant $c_{crit}$ is not yet known, as the localized black hole solutions have not yet been constructed in the correct context to be embedded in the supergravity dual.  The line
 $r_x^2 = c_{crit} \, r_\tau$ represents a first order thermal phase transition between the uniform and localized solutions, with uniform favored for higher temperature $r_x^2 > c_{crit} \, r_\tau$ and localized favored for lower temperature $r_x^2 < c_{crit} \, r_\tau$.\footnote{The region $r_x^2 < \alpha\, r_\tau$ for $2.29 < \alpha < c_{crit}$ is the region where the localized solution dominates the canonical ensemble, but the uniform phase could in principle be constructed as a metastable supercooled state (although here we will only be concerned with equilibrium thermodynamics).} 
We term this the GL phase transition and emphasize that this is distinct from the GL dynamical instability. 
 Whilst there is a non-uniform black hole solution it is never thermally dominant. For reviews on the GL dynamical instability, phase transition and uniform, non-uniform and localized black hole solutions see \cite{Kol:2004ww,Harmark:2005pp,Harmark:2007md}.

According to the duality hypothesis, a Polyakov loop about the time/space circle in the Euclidean SYM is computed in the leading large $N$ limit by considering whether a two dimensional minimal area surface (the classical string worldsheet) that asymptotically wraps the time/space circle exists. If the time/space circle is contractible in the interior of the gravity solution, a minimal area solution for the string worldsheet will exist and then the correspondence states that $P_{\tau/x} \sim O(1)$. However, if the circle is not contractible, there cannot exist a minimal surface that gives a finite action for the string worldsheet, and the correspondence states that $P_{\tau/x} \sim O(1/N)$ and hence $P_{\tau/x} = 0$ in the large $N$ limit. It is a standard result of Euclidean gravity that black hole solutions have contractible time circles in the interior of the solution, and in fact the time circle contracts precisely at the horizon. The contractability of the spatial circle however depends on the type of black hole. In the IIB supergravity solution the space circle is non-contractible. The IIA uniform (and non-uniform) solutions have non-contractible space circles, whereas the localized solution has a contractible circle. In fact the eigenvalues of the SYM spatial Polyakov loop (which are phases, and hence live on a circle) are thought to correspond to the positions of these $D0$-branes on the space circle in the IIA dual. Hence the GL phase transition can physically be thought of as a thermal instability 
associated with the clumping of $D0$-branes, breaking the $U(1)$ circle translation symmetry. In the large $N$ SYM this symmetry breaking is the spontaneous breaking of center symmetry $Z_N$, where for large $N$, $U(1)\simeq Z_N$.

Let us summarize our predictions for the large torus. We learn that in the IIB regime,  $1 \ll r_\tau \ll r_x^2$, we expect $P_\tau \ne 0$ but $P_x = 0$. In the IIA regime, where $1 \ll r_\tau$ and $r_x^{4/3} \ll r_\tau$, we have $P_\tau \ne 0$, and $P_x \ne 0$ for $r_x^2 \leq c_{crit} \, r_\tau$ and $P_x = 0$ for $r_x^2 > c_{crit} \, r_\tau$, with $c_{crit}$ an order one constant with $c_{crit} > 2.29$. We note that in the regime where both IIA and IIB apply, they give consistent results. Thus in the large torus, supergravity regimes, the SYM is always deconfined in the time direction, and there is a first order deconfinement/confinement transition in the space direction at $r_x^2 = c_{crit} \, r_\tau$.

\subsection{Dimensional reduction}

Consider the toy model scalar theory defined on the 2-torus,
\begin{eqnarray}
S = \frac{1}{\lambda} \int_{T^2} d\tau dx \left(   ( \partial_\mu \phi )^2 + \phi^4  \right)
\end{eqnarray}
First we change to angular coordinates $\theta_\tau = \tau/\beta $ and $\theta_x = x/R$ with unit radius, so $\theta_{\tau,x} \sim \theta_{\tau,x}+1$, and then define the dimensionless scalar variable $\tilde{\phi} = \left( \beta R / \lambda \right)^{1/4} \phi$. The action can now be written as,
\begin{eqnarray}
S = \int_0^{2 \pi} d\theta_\tau d\theta_x \left( \tilde{\phi}^4 + \sqrt{\frac{r_x}{r_\tau^3}} ( \partial_{\theta_\tau} \tilde{\phi} )^2  +  \sqrt{\frac{r_\tau}{r_x^3}} ( \partial_{\theta_x} \tilde{\phi} )^2 \right)
\end{eqnarray}
and we see that the dimensionless couplings $r_x/r_\tau^3$ and $r_\tau/r_x^3$ determine the masses of the non-constant modes of the field $\phi$ on the torus.  There are three interesting limits. When $r_x \sim r_\tau \ll1$, then the non-constant modes of the scalar become very massive and hence weakly coupled and one may integrate these out to arrive simply at the quartic integral governing the constant modes. If only $1 \ll r_x/r_\tau^3$ then the non-constant modes on the time circle are weakly coupled and one may integrate these out to obtain the dimensional reduction which now lives only on the space circle. Likewise, if $1 \ll r_\tau/r_x^3$, one may dimensionally reduce to obtain a theory only on the time circle.

\FIGURE[h]{
 \centerline{ 
 \includegraphics[width=.6\textwidth]{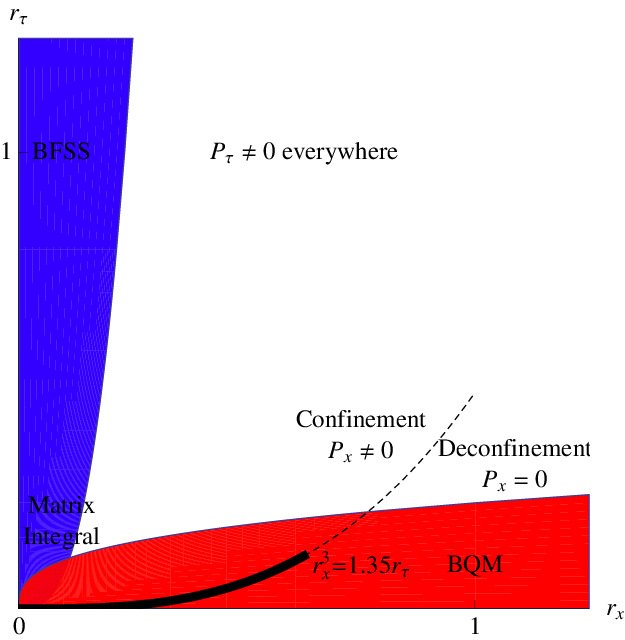}
 }
         \caption{Figure showing the regions of coupling space where the SYM may be dimensionally reduced on the time and/or space circles. The blue region indicates the region where reduction on the space circle gives a good approximation, yielding a supersymmetric quantum mechanics theory, the BFSS model. The red region indicates where reduction on the space circle to a bosonic quantum mechanics (BQM) is a good approximation. This latter reduction predicts a large $N$ deconfinement phase transition in the spatial Polyakov loop for $r_x^3 = 1.35 r_{\tau}$ and this curve is shown.}
   \label{fig:dimred}
}

The structure of this toy example is such that precisely the same phenomena occurs with the full SYM on a 2-torus, as discussed in \cite{Aharony:2005ew}. One difference is that due to the anti-periodic boundary conditions on the time circle, the Fourier decomposition of the fermions contain only non-constant modes in the time direction. Another difference is that under a reduction, the constant component of the gauge field in the direction of reduction yields a scalar field, similar to the scalars $\phi^I$, in the reduced theory. This scalar in the reduced theory corresponds to the Polyakov loop about the cycle that has been reduced on. Since the expectation value of the eigenvalue distribution of the scalar in the reduced theory will have a non trivial profile, this implies that center symmetry is broken in the Polyakov loop about the reduced cycle \footnote{This is to be contrasted with the Eguchi-Kawai reduction \cite{Eguchi:1982nm} where quite the opposite occurs; one can only reduce on a direction if center symmetry is unbroken.}.

There are again 3 regimes. For $r_x \sim r_\tau \ll 1$ one may reduce on both time and space to just give the zero modes of the theory, and arrive at a bosonic Yang-Mills matrix integral, since in reducing on the time circle one loses the fermions which have no zero modes. Such a reduction indicates that in this limit, the two dimensional SYM should have $P_\tau , P_x \ne 0$.

For $r_x^3 \ll r_\tau$ the theory may be dimensionally reduced on the space circle to give the thermal supersymmetric matrix quantum mechanics living on the time circle with radius $\beta$. The spatial Polyakov loop is 
then given in terms of one of the 9 scalars of the BFSS model, and since these scalars have localized eigenvalues, the two dimensional SYM should be deconfined in the space direction with $P_x \ne 0$.  This theory is precisely the BFSS theory  \cite{Banks:1996vh}, and recently this has been numerically simulated in the 't Hooft limit \cite{Catterall:2007fp,Anagnostopoulos:2007fw,Catterall:2008yz}, and indeed, the results obtained are consistent with the theory always being deconfined, so $P_{\tau} \ne 0$. The coupling of this quantum mechanics is given by $r_\tau / ( r_x )^{1/3}$ and when this is large we know from our arguments above that we are in a regime where a dual IIA supergravity description exists, and the dynamics is given by the localized black hole solution which is indeed consistent with $P_\tau, P_x \ne 0$. 

For $r_\tau^3 \ll r_x$ one may again perform a dimensional reduction, now on the time circle. Thus in the two dimensional theory we expect $P_\tau \ne 0$. Since there are no fermion zero modes on the time circle, the resulting one dimensional theory is a bosonic quantum mechanics (BQM) defined on a circle radius $R$ and with dimensionless coupling $r_x^3 / r_\tau$. Numerical \cite{Aharony:2004ig,Aharony:2005ew,Kawahara:2007fn} and analytic  study \cite{Mandal:2009vz} indicates that this theory has a large $N$ confinement/deconfinement transition at $r_x^3 / r_\tau \simeq 1.35$ of second order. There is also thought to be a third order Gross-Witten-Wadia \cite{GW, Wadia} transition very nearby at $r_x^3 / r_\tau \simeq 1.49$ \cite{Kawahara:2007fn,Mandal:2009vz}.

\subsection{Expectations for large $N$ phase diagram}

We conclude by putting together the above discussions. The simplest picture is then that the Gregory-Laflamme first order phase transition, $r_x^2 = c_{crit} \, r_\tau$ for $1 \ll r_\tau$ (recall $c_{crit} > 2.29$), and the second order transition $r_\tau^3 \ll r_x$ and $r_x^3 = 1.35 \, r_\tau$ in the time reduced BQM are two ends of the same spatial Polyakov loop confinement/deconfinement phase transition line. At some point in-between the order presumably changes, and here the new third order Gross-Witten-Wadia  phase transition emerges although this is not measured by centre symmetry breaking, but by more detailed information about the spatial Polyakov loop eigenvalue distribution. It is interesting \cite{Kol:2002xz,Aharony:2004ig,Harmark:2004ws} that the new phase at small $r_x$ also exists for $1 \ll r_x$ in the form of  non-uniform IIA black strings, but unlike at weak coupling, these are never thermally dominant in the IIA supergravity region. In figure \ref{fig:phase} we summarize the expected phase diagram for the spatial confinement/deconfinement transition.

\FIGURE[h]{
 \centerline{ 
 \includegraphics[width=.6\textwidth]{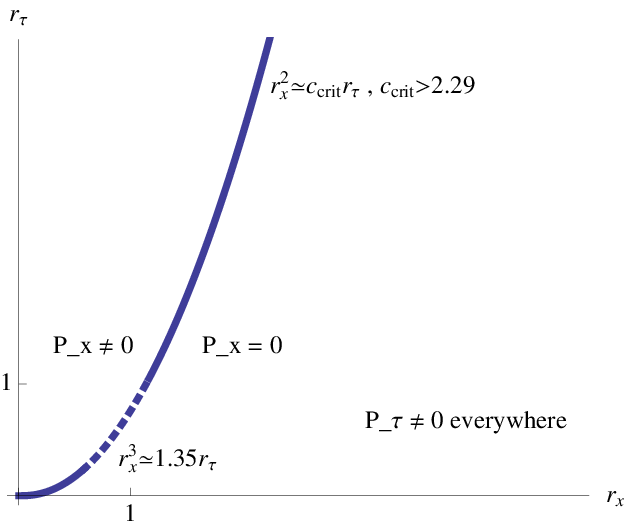}
 }
         \caption{Cartoon of the expected large $N$, spatial Polyakov loop deconfinement transition line in coupling space. Pictured is the simplest possibility, namely that the spatial deconfinement transition interpolates between the strong coupling Gregory-Laflamme transition parametric behavior $r_x^2 \sim r_\tau$, and the high temperature reduction deconfinement transition behavior $r_x^3 = 1.35 r_\tau$.}
   \label{fig:phase}
}

%
\section{Supersymmetric lattices for super Yang-Mills}
\label{sec:latticeact}
%

In recent years there has been significant progress in the formulation
of lattice theories which possess exact supersymmetry - see \cite{Catterall:2009it,Catterall:2010nd}
and references therein. The chief advantage of these approaches is that
they significantly reduce, or in many cases eliminate, the need to fine tune
the couplings in the lattice theories to approach the target continuum
supersymmetric field theory as the lattice spacing is sent to zero.
The case of ${\cal N}=4$ super Yang-Mills is particularly interesting and
the corresponding lattice theory was first constructed using
orbifold methods in \cite{Kaplan:2005ta}. Subsequently, it was realized that the
same theory could be obtained using a carefully chosen discretization of
a topologically twisted version of the continuum theory \cite{Unsal:2006qp,Catterall:2007kn}.

The continuum twist of $\cN=4$ that is the
starting point of the twisted lattice construction was
first written down by Marcus in 1995 \cite{Marcus:1995mq} although it now plays an 
important role
in the Geometric-Langlands program and is hence sometimes called
the GL-twist \cite{Kapustin:2006pk}. This four dimensional twisted theory 
is most compactly expressed
as the dimensional reduction of a five dimensional theory in which the
ten (one four component gauge field and six scalars) bosonic fields
are realized as the components of a complexified five dimensional
gauge field $\cA_a,\,a=1\ldots 5$
while the 16 single component twisted fermionic degrees of
freedom are realized as the 16 components of a 
\KD field $(\eta,\psi_a,\chi_{ab})$ \cite{Catterall:2007kn}. 
The appearance of a scalar fermion after twisting is crucial -- it implies the
existence of a nilpotent supersymmetry which will be preserved in the
lattice theory. Its action on the continuum fields\footnote{We assume
an anti-hermitian basis for all fields which take their
values in the adjoint representation of the $SU(N)$ gauge group.} is given by
\begin{eqnarray*}
\cQ\; \cA_a&=&\psi_a\nonumber\\
\cQ\; \psi_a&=&0\nonumber\\
\cQ\; \cAb_a&=&0\nonumber\\
\cQ\; \chi_{ab}&=&-\cFb_{ab}\nonumber\\
\cQ\; \eta&=&d\nonumber\\
\cQ\; d&=&0
\label{Qsusy}
\end{eqnarray*}
The site field $d$ is a non-dynamical field that is included to
close the $\cQ$-algebra and is subsequently integrated out of the
final lattice action. 
Furthermore, the action of
this theory can be written as the sum of two terms;
a $\cQ$-exact piece of the form
\beq
S= \frac{1}{4 g_{YM}^2} \cQ
\int\Tr\left(\chi_{ab}\cF_{ab}+\eta [ \cDb_a,\cD_a ]-\frac{1}{2}\eta
d\right)\label{2daction}\eeq
and an additional
$\cQ$-closed term 
\beq
S_{\rm closed}=-\frac{1}{32 g_{YM}^2}\int \Tr 
\epsilon_{mnpqr}\chi_{qr}\cDb_p\chi_{mn}  \, .
\label{closed} 
\eeq  
The supersymmetric invariance of this term then relies on the
Bianchi identity 
\beq\epsilon_{mnpqr}\cDb_p\cFb_{qr}=0\eeq 
Discretization of this theory proceeds straightforwardly;
(Complex) continuum
gauge fields are represented as complexified Wilson gauge links
$\cU_\mu(x)=e^{\cA_\mu(x)}$ living on links 
$e_\mu,\,\mu=1\ldots 4$ of a hypercubic lattice.\footnote{A better choice in four dimensions is the $A_4^*$ lattice which retains
a higher point group symmetry than the hypercubic lattice. See \cite{Catterall:2009it} for
details. It is not necessary for two dimensions and indeed would complicate
the calculation of Polyakov lines.} 
The field $\cU_5$ is an exception to this and is placed on the
body diagonal of the hypercube corresponding to a relative position vector $e_5=(-1,-1,-1,-1)$. Notice that $\sum_{a=1}^5 e_a=0$
which is important to show gauge invariance of the lattice action.
These fields transform in the usual way under $U(N)$ lattice 
gauge transformations eg:
\beq
\cU_a(x)\to G(x)\cU_a(x)G^\dagger(x+e_a)\eeq
Supersymmetric invariance
then implies that $\psi_a(x)$ live on the corresponding links
and transform identically to $\cU_a(x)$. 
The scalar fermion $\eta(x)$ is clearly most naturally associated with
a site and transforms accordingly
\beq \eta(x)\to G(x)\eta(x)G^\dagger(x)\eeq
The field $\chi_{ab}$ is slightly more difficult. Naturally as a 2-form
it should be associated with a plaquette. In practice we introduce diagonal
links running through the center of the plaquette and
corresponding to the vector $e_{ab}=e_a+e_b$
and choose $\chi_{ab}(x)$
to lie {\it with opposite orientation} along those diagonal links. This
choice of orientation will again be necessary to ensure gauge
invariance. The scalar lattice supersymmetry transformation is identical
to that in the continuum after the replacement $\cA_a\to\cU_a$. 
Most importantly it remains nilpotent which means that we can guarantee
invariance of the $\cQ$-exact
part of the lattice action by replacing the continuum fields by their
lattice counterparts.
Of course to do so necessarily requires
a prescription for replacing continuum derivative operators
by gauge covariant finite difference operators. 
The following expressions are used:
\begin{eqnarray}
\cD^{(+)}_a f_b&=&
\cU_a(x)f_b(x+e_a)-f_b(x)\cU_a(x+e_b)\\
\cDb^{(-)}_a f_a&=&f_a(x)\cUb_a(x)-\cUb_a(x-e_a)f_a(x-e_a)
\end{eqnarray}
Notice that this definition reduces to the usual adjoint covariant derivative
in the naive continuum limit and furthermore guarantees that the
resultant expressions transform covariantly under lattice gauge
transformation.
The lattice field strength is then given by 
the gauged forward difference $\cF_{ab}=\cD^{(+)}_a \cU_b$
and is automatically antisymmetric in its indices.
Furthermore it transforms like
a lattice 2-form or plaquette and hence
yields a gauge invariant loop on the lattice when contracted
with the plaquette fermion $\chi_{ab}$.
Similarly the covariant discrete divergence appearing in $\cDb_a^{(-)} \cU_a$
transforms as a 0-form or site field and hence can be contracted with
the site field $\eta$ to yield a gauge invariant expression.

This use of forward and backward difference operators guarantees that the
solutions of the theory map one-to-one with the solutions of the continuum
theory and hence fermion doubling problems are evaded \cite{Rabin:1981qj}.
Indeed, by introducing a lattice with half the lattice spacing one can
map this \KD fermion action into the action for staggered fermions \cite{Banks:1982iq}. 
While the supersymmetric invariance of the $\cQ$-exact term is manifest
in the lattice theory it is not clear how to discretize the
continuum $\cQ$-closed term. Remarkably, it is possible to discretize (\ref{closed})
in such a way that it is indeed exactly invariant under the twisted
supersymmetry
\beq
S_{\rm closed}=-\frac{\kappa}{8}\sum_{\bx}\Tr 
\epsilon_{mnpqr}\chi_{qr}(\bx+e_m+e_n+e_p)
\cDb^{(-)}_p\chi_{mn}(\bx+e_p)\eeq
with $\kappa$ the lattice coupling, and can be seen to be supersymmetric since the lattice field
strength satisfies an exact Bianchi identity \cite{Aratyn:1984bd}
\beq
\epsilon_{mnpqr}\cDb^{(+)}_p\cFb_{qr}=0 \, .
\eeq
Putting all these elements together we arrive at the supersymmetric
lattice action \cite{Catterall:2009it}
\beq
S=S_{\rm closed}+\kappa \sum_{\bx}\Tr \left(\cF^{\dagger}_{ab}\cF_{ab}+
\frac{1}{2}\left(\cDb^{(-)}_a \cU_a\right)^2-
\chi_{ab}\cD^{(+)}_{\left[a\right.}\psi_{\left.b\right]}-
\eta \cDb^{(-)}_a\psi_a\right)
\label{4action}
\eeq
where we have taken the $\cQ$-variation and integrated out the
auxiliary field $d$. To reiterate; this action is gauge invariant, free of
doublers and possesses the one exact supersymmetry given in (\ref{Qsusy}).

Finally to obtain a two dimensional theory we perform a simple dimensional reduction along two lattice directions using periodic boundary conditions. The
resultant lattice action corresponds in the 
naive continuum limit to the target
$\cQ=16$ YM theory in two dimensions. In this limit its exact
supersymmetry is enhanced to correspond to 4 continuum supercharges
corresponding to the four scalar fermions that now appear in the dimensionally
reduced theory \cite{Catterall:2009it}.

We will be interested
in the continuum limit of this theory, given in equation~\eqref{eq:SYM}, in the
large $N$ limit with
't Hooft coupling $\lambda=N g_{YM}^2$. The 
lattice theory is then governed by the coupling
$\kappa=\frac{1}{4 g_{YM}^2} \frac{T^2}{\beta^2} = \frac{N T^2}{4 r_\tau^2}$ 
where $T$ denotes the number of
lattice sites in the
 temporal directions.

We have used periodic boundary conditions for the fields on the 
remaining spatial circle and anti-periodic boundary conditions for fermions 
in the temporal direction
in order to access the thermal theory. Simulations were carried out using
the RHMC algorithm which is described in detail in \cite{Catterall:2008dv}.
It has been
shown that the existence of a noncompact moduli space in the theory renders the thermal partition function divergent
\cite{Catterall:2009xn}. In order to regulate this divergence we have additionally
introduced a mass term for the scalar fields 
appearing in the lattice action with a dimensionless mass parameter $m=
m_{\rm phys} \beta$, 
\beq
S_m= m^2 \, \kappa \sum_x \left[\cU_\mu^\dagger\cU_\mu+
\left(\cU_\mu^\dagger\cU_\mu\right)^{-1}-2\right] \, .
\eeq
The form of this term is effective at suppressing arbitrarily
large fluctuations of the exponentiated scalar fields
and reduces to a simple mass term for small fluctuations characterizing
the continuum limit. Notice that
this infrared regulator
term breaks supersymmetry softly and lifts the quantum moduli space of the theory. We have performed our simulations for a range of the parameter
$m$ in order to allow for an extrapolation $m\to 0$. 

%
\section{Simulation results}
%

\FIGURE[h]{
 \centerline{ 
 \includegraphics[width=.8\textwidth]{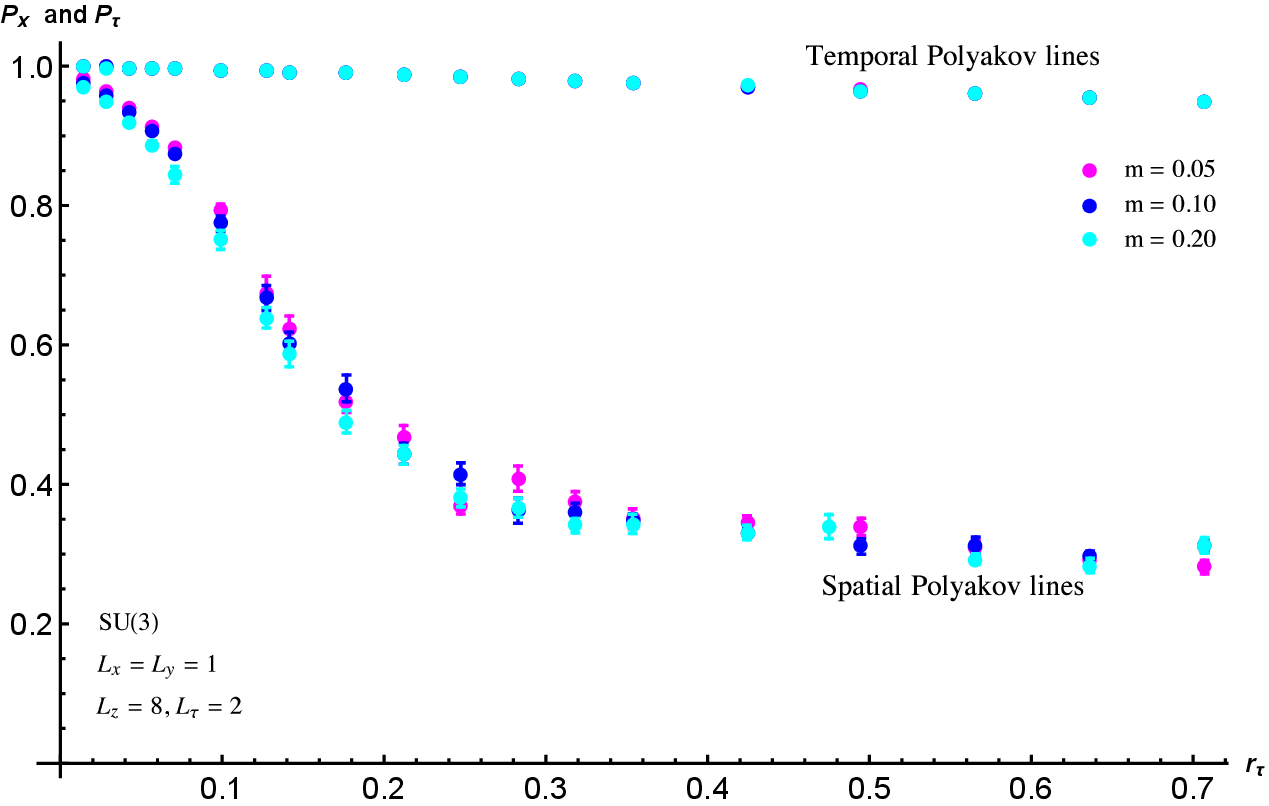}}
  \centerline{ 
  \includegraphics[width=.8\textwidth]{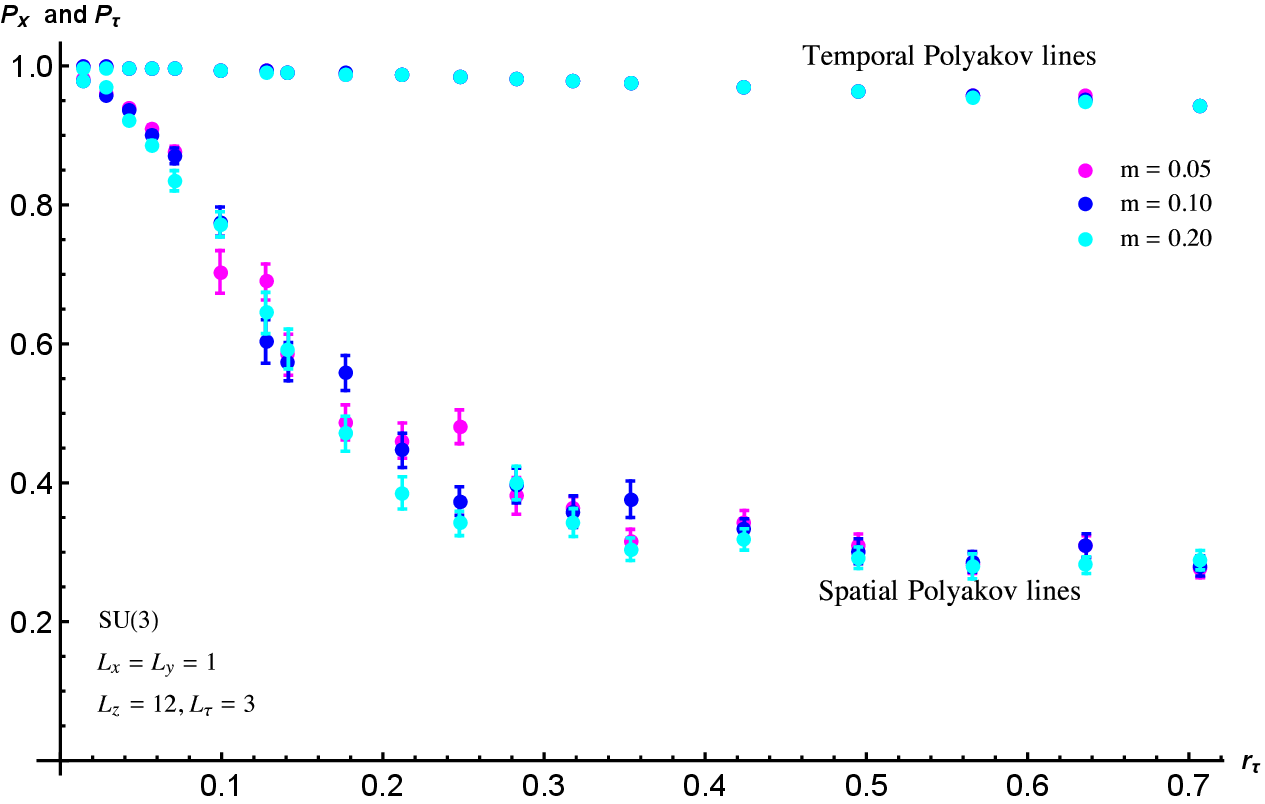}}
        \caption{Spatial and temporal Polyakov lines ($P_x$ and $P_{\tau}$) against dimensionless time circle radius $r_\tau$ for maximally supersymmetric $SU(3)$ Yang-Mills on $2 \times 8$ and $3 \times 12$ lattices using different values of the infrared regulator m.}
   \label{fig:transition}
}

In this section we present our numerical results. We have focused on
the Polyakov lines for both the thermal and spatial circle. These are
defined in the usual way
\beq
P_x = \frac{1}{N} \Big{\langle} \Big|\Tr \Pi_{a_x=0}^{L-1} U_{a_x}\Big| \Big{\rangle},~~P_\tau = \frac{1}{N} \Big{\langle} \Big|\Tr \Pi_{a_\tau=0}^{T-1} U_{a_\tau}\Big| \Big{\rangle}~,
\eeq
where we have extracted the unitary piece of the complexified link
$\cU_\mu$ to compute these expressions.
We have evaluated the spatial and temporal Polyakov lines as a function
of $r_\tau$ for two different lattices with the same aspect ratio, a $2 \times 8$ lattice and a $3 \times 12$ lattice, for $N=3$ and with values of the
infrared regulator $m = 0.05, 0.10$ and $0.20$. 
The use of two different lattices with the same
aspect ratio allows us to test for and quantify
finite lattice spacing effects.
We have performed simulations for values of the dimensionless time circle radius in the range $0.01 \leq  r_\tau \leq 0.7$. 
The results are shown in
figure~\ref{fig:transition}.

Notice that the temporal Polyakov
remains close to unity over a wide range of $r_\tau$. This indicates the theory is (temporally) deconfined and is consistent
with expectations for the limits discussed in section \ref{sec:theory} - the asymmetric torus limits, and the strong coupling regions where there is a dual supergravity description in terms of black holes. However,
the spatial Polyakov line has a different behavior
taking values close to unity for small $r_\tau$ while
falling rapidly to
plateau at much smaller values for large $r_\tau$. It is tempting to
see the rather rapid crossover around $r_\tau \sim 0.15$ as
a signal for a would be thermal phase transition
as the number of
colors is increased. 
This conjecture is seen to be consistent with the data; in figure~\ref{fig:color} we show the Polyakov lines for
$N=2,3,4$ on $2\times 8$ lattices
as a function of $r_\tau$. The plateau evident at large $r_\tau$ falls
with increasing $N$ and the crossover sharpens. This is consistent with
the system developing 
a sharp phase transition in the large $N$ limit. 

\FIGURE[t]{
\includegraphics[width=.8\textwidth]{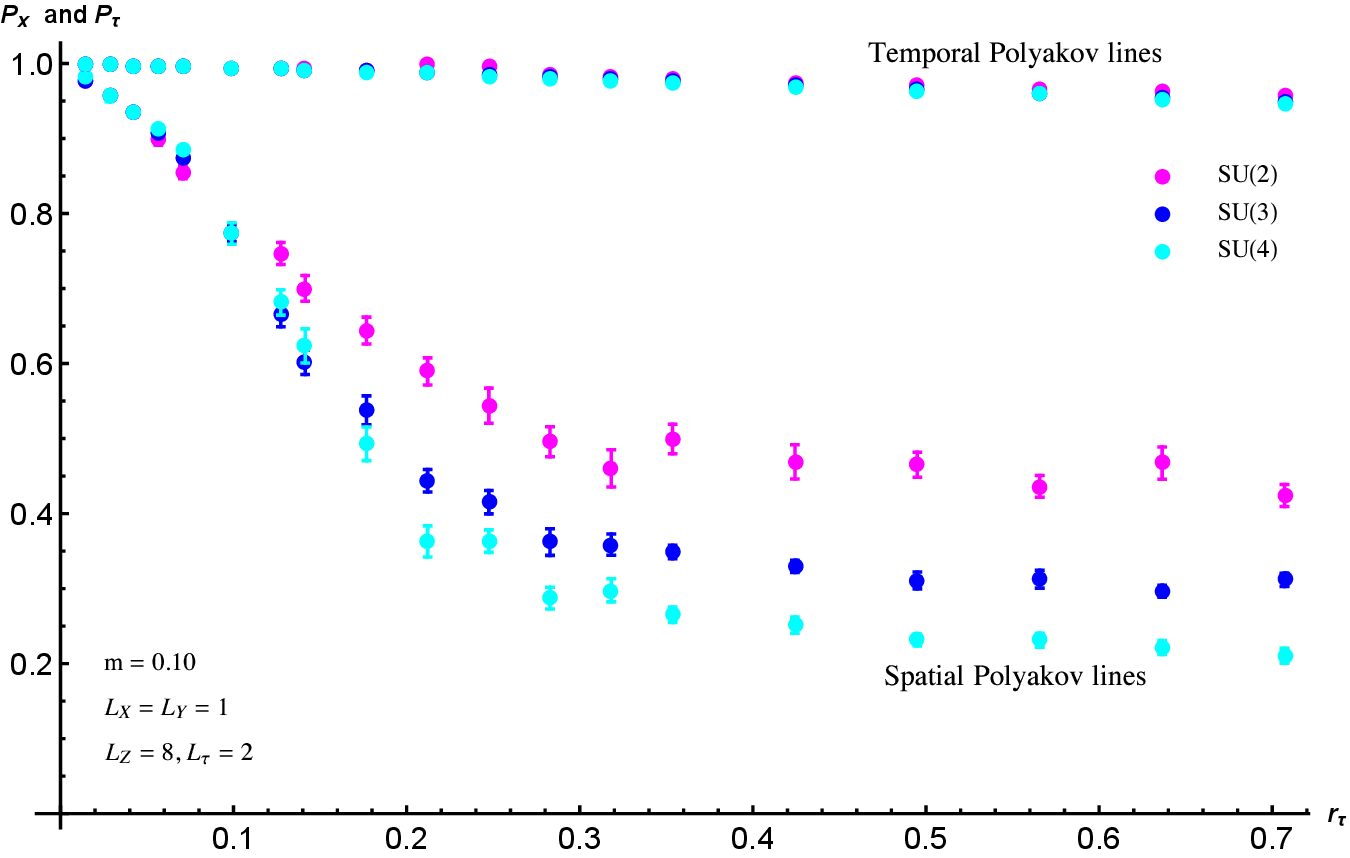}
\caption{\label{fig:color} \small{Plot of the absolute values of the spatial and temporal Polyakov lines ($P_x$ and $P_{\tau}$) against the dimensionless time circle radius $r_\tau$ for maximally supersymmetric $SU(N)$ Yang-Mills on a $2 \times 8$ lattice for $N = 2, 3, 4$, using the value of the infrared regulator m = 0.10.}}
}

In the data shown here we see very little dependence
of our results on the scalar mass. Indeed for the length of Monte Carlo we were able to perform it appears that $m$ can be set
to zero for 
$r_\tau<1.5$ 
without fear of encountering the thermal
divergence discussed in \cite{Catterall:2009xn}.
This stability in the scalar sector can be seen in figure~\ref{fig:scalar-eigenvalues}
which shows the Monte Carlo time series for the eigenvalues of $U_\mu^\dagger U_\mu\sim e^{2\phi}$
at two different $r_\tau$'s with dimensionless mass parameter m = 0.05 and gauge group $SU(3)$. There is no evidence of
a divergence over thousands of Monte Carlo sweeps. Furthermore, one sees that the eigenvalues of the scalar fields
(rendered dimensionless using the lattice spacing) cluster with small separation for this range of $r_\tau$.\footnote{We thank
Masanori Hanada for pointing out the significance of this result which differs from the situation reported in
\cite{Hanada:2009hq}}

\FIGURE[h]{
 \centerline{
 \put(213,36){${}_{\frac{}{\sqrt{2}}}$}
 \includegraphics[width=.6\textwidth]{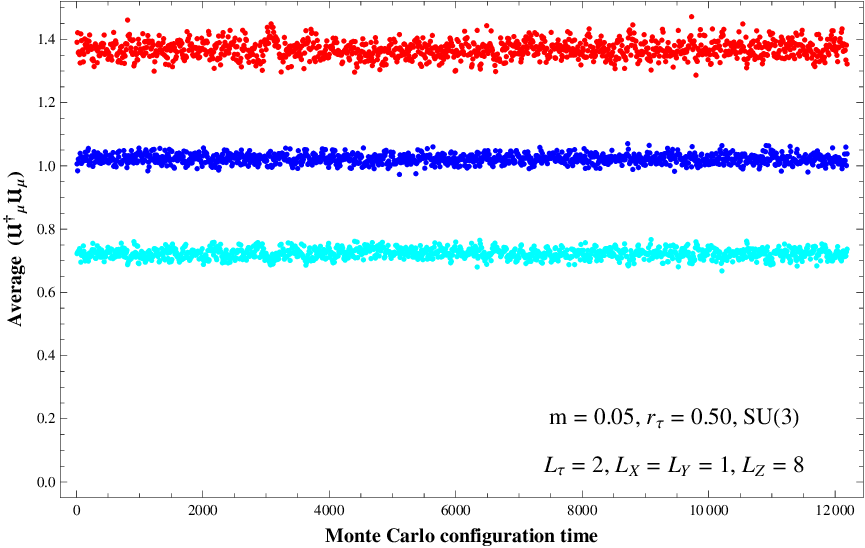}
}
    \centerline{ 
     \put(219,33){${}_{\frac{}{\sqrt{2}}}$}
     \includegraphics[width=.6\textwidth]{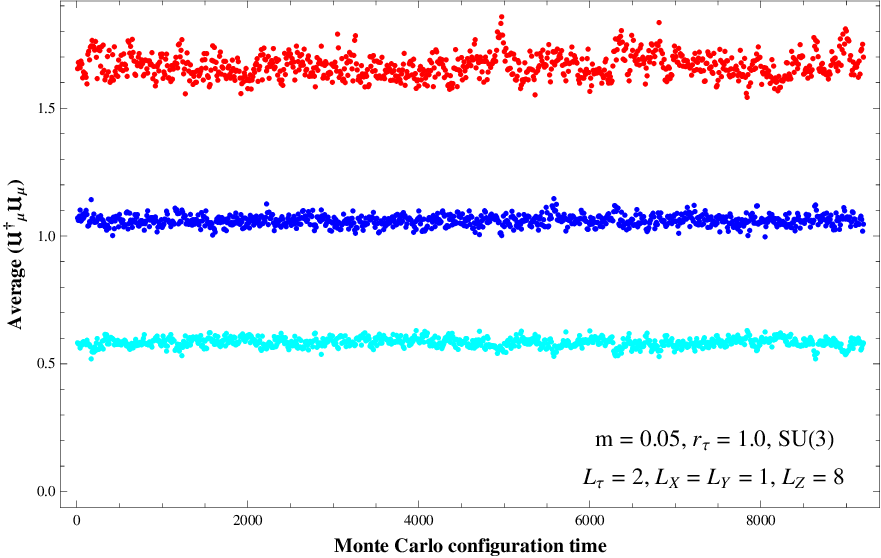}
}
        \caption{Plots of the average scalar eigenvalues against Monte Carlo configuration time step, for $N=3$ on a $2 \times 8$ lattice with 
        $\sqrt{2} \, r_\tau = 0.5$ and  $1.0$. 
        Note that the spread between eigenvalues reduces as $r_\tau$ is decreased. We have used the dimensionless mass parameter m = 0.05.}
   \label{fig:scalar-eigenvalues}
}
We have however observed that
the $m=0$ model does exhibit the same thermal instability
observed in the case of supersymmetric quantum mechanics
for sufficiently low temperature $r_\tau >>1$ in agreement
with the general arguments given in \cite{Catterall:2009xn}.

\FIGURE[h]{
\includegraphics[width=.48\textwidth]{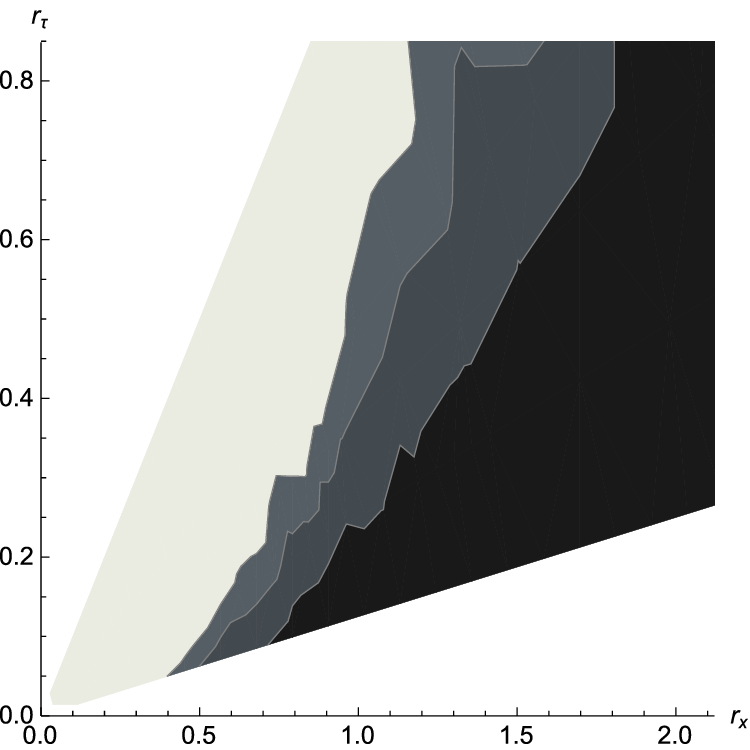} \includegraphics[width=.48\textwidth]{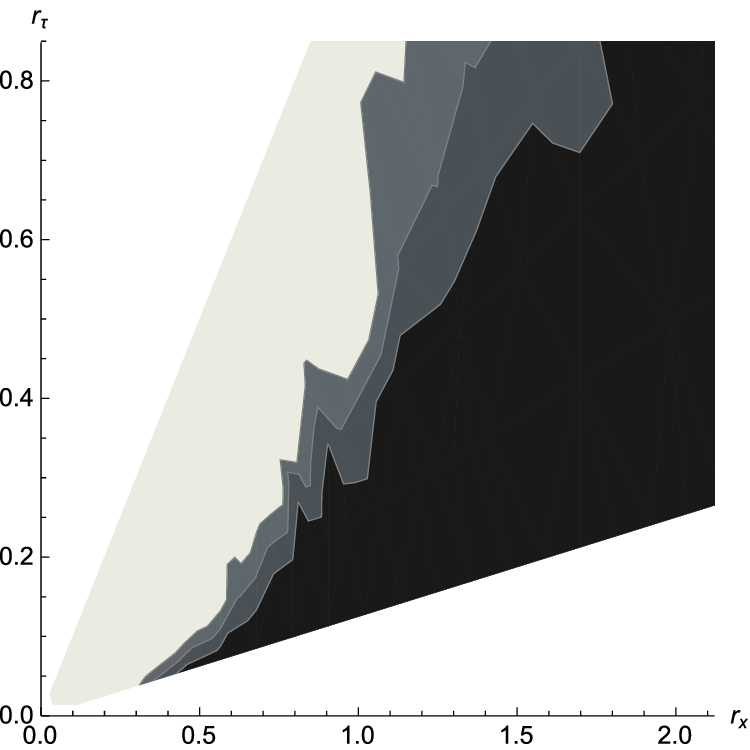}
\caption{\label{fig:contour} \small{Plot of contours of the expectation of the spatial Polyakov line $P_x$ over the $r_x, r_\tau$ plane. The left frame shows $SU(3)$, and the right $SU(4)$. The three contours plotted are $0.4, 0.5, 0.6$, and the simulation data collates and interpolates runs made on lattices $2 \times 16$, $2 \times 8$, $3 \times 8$, $4 \times 4$ and $4 \times 8$ therefore giving a variety of aspect ratios $r_\tau/r_x$.}}
}

Putting together several lattice aspect ratios for $N = 3, 4$ we can plot the spatial Polyakov loop as a function of $r_s$ and $r_\tau$ where data is available. This is done in figure~\ref{fig:contour}. The three contours $P_x = 0.4, 0.5, 0.6$ are shown. We see that the contours for $SU(4)$ are closer together than those for $SU(3)$, as we expect for a large $N$ transition.
From this data we can try to assess where the large $N$ transition in $P_x$ may occur. In the detailed studies of the dimensionally reduced bosonic quantum mechanics \cite{Aharony:2004ig}, it was found that the large $N$ transition occurred very close to $P_x \simeq 0.5$. Thus from the contours of the $SU(3)$ and $SU(4)$ data we could take the $P_x = 0.5$ curves to give an estimate for the large $N$ phase transition line. 
Another estimate is to plot the function $f_n \equiv P_x(SU(n)) - P_x(SU(n-1))$ which measures the difference between the Polyakov lines for $SU(n)$ and $SU(n-1)$. At strong coupling where we expect the large $N$ transition is first order, the simplest situation is to have $f_n <  0$ in the confined region (where $P_x = 0$ for $N \rightarrow \infty$), and correspondingly $f_n > 0$ in the deconfined region as $n \rightarrow \infty$. Then plotting the boundary of the positive (or negative) region of $f_4$ calculated from our data also gives an estimate of the critical line. Neither method can give a precise determination, and they should not be considered as a replacement for calculations at larger $N$ than we have been able to reach here. However, in the absence of such large $N$ data we plot the $P_x = 0.5$ contours for $SU(4)$ and $SU(3)$ in figure~\ref{fig:contour2}, and in addition the region where $f_4$ is positive. We note that the $SU(3)$ and $SU(4)$ $P_x = 0.5$ contours are remarkably consistent with each other, which provides evidence that they are indeed a reasonable approximation to the large $N$ transition curve. Whilst the $f_4$ data is rather noisy, and hence the positive $f_4$ region has `holes' in it, the function is positive only to left of the $P_x = 0.5$ curves, and furthermore, extends right up to these curves. The curve 
$r_x^2 = 2.29 \, r_\tau$, which gives the GL instability boundary,
 is plotted on this graph and matches the contours $P_x = 0.5$ and the boundary of the positive $f_4$ region very well in the strong coupling region. We take this to indicate that the gravity prediction for the parametric behavior $r_x^2 = c_{crit} r_\tau$ is consistent with our data, and we have estimated 
a $c_{crit}$ very close to the value for the GL instability, ie. $c_{crit} \simeq 2.29$. Presumably it is a little larger, as the GL phase transition is bounded by the instability, but we cannot distinguish the small difference with our results.
Furthermore we see that the contours $P_x = 0.5$ also appear to be consistent with the high temperature prediction $r_x^3 = 1.35 r_\tau$ as well.

\FIGURE[t]{
\includegraphics[width=.7\textwidth]{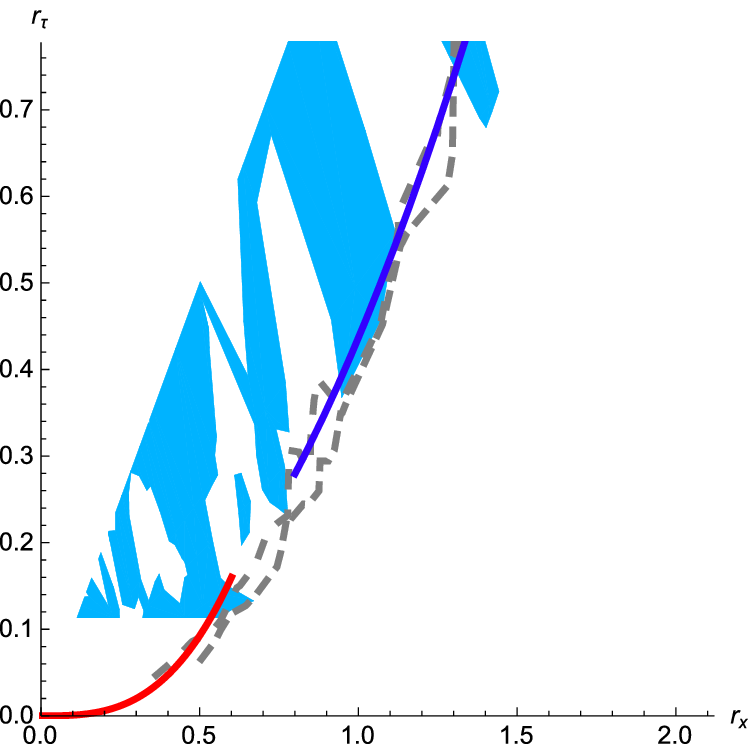}
\caption{\label{fig:contour2} \small{Plot showing a superposition of the $P_x = 0.5$ contours for $SU(3)$ and $SU(4)$ as dashed black lines. Also shown is the region (blue) where the $SU(4)$ loop $P_x$ is greater than the $SU(3)$ loop, which is expected to estimate the large $N$ deconfined region for a first order transition (which gravity suggests at strong coupling). `Holes' in this blue region are due to statistical errors. We see the boundary of this region (ignoring `holes') matches well the $P_x = 0.5$ contours, and represents our guess for where the large $N$ transition resides. This figure should be compared to the previous figure \ref{fig:phase} giving a sketch of the expected phase structure. Plotted on the figure is the high temperature prediction for the transition ($r_x^3 = 1.35 r_\tau$, red curve). We note that the estimated large $N$ transition curve fits well both this high temperature prediction and also the strong coupling dual gravity predicted parametric behavior  $r_x^2 = c_{crit} r_\tau$. Our data suggests 
$c_{crit}$ for the GL transition is very close to the value $2.29$ for the GL instability (plotted as blue curve).
}}
}

The value of the ratio $\alpha \equiv c_{crit}/2.29$ gives the ratio of the GL thermal phase transition temperature to the GL dynamical instability temperature (the minimum temperature to which uniform strings can be supercooled), so $\alpha = T_{GL\; phase}/T_{GL\; instab}$. Whilst the GL instability temperature is known \cite{Aharony:2004ig} (corresponding to the behavior $r_x^2 = 2.29 r_\tau$ at strong coupling), the GL phase transition temperature is not known in the gravity theory as the localized solutions have not been constructed. 
In fact the near extremal D0-charged black holes are simply related to vacuum solutions of pure gravity with $\mathbb{R}^{1,8}\times S^1$ asymptotics \cite{Aharony:2004ig,Harmark:2004ws}. Such localized black hole solutions have been constructed for asymptotics $\mathbb{R}^{1,3}\times S^1$  and $\mathbb{R}^{1,4}\times S^1$, using numerical techniques \cite{Wiseman:2002zc,Kudoh:2003ki,Headrick:2009pv}. Extending these methods to the case of interest here, $\mathbb{R}^{1,8}\times S^1$, is obviously an interesting future direction. It is worth emphasizing that whilst finding localized solutions in the gravity theory only involves solving the classical Einstein equations, in practice even phrasing the Einstein equations in a manner amenable to numerical solution has presented a challenge \cite{Headrick:2009pv} and then solving the resulting coupled partial differential equations is a serious numerical undertaking.\footnote{Such solutions can be constructed perturbatively \cite{Harmark:2003yz,Gorbonos:2004uc,Gorbonos:2005px,Karasik:2004ds} in a small radius limit (compared to the circle size) but the GL phase transition occurs for black holes with radius of order the circle size, and hence it is unclear how accurate perturbative methods are for a prediction of $T_{GL\; phase}/T_{GL\; instab}$.}
Our lattice estimation 
$\alpha \simeq 1.0$ 
provides a prediction for the thermal behavior of the gravity solutions. To our knowledge, this is the first time a prediction about the properties of non-trivial classical gravity solutions has been made from the Yang-Mills side of a holographic correspondence.

%
\section{Conclusions}
%

We have used a supersymmetric lattice action to study the strongly coupled
dynamics of two dimensional maximally supersymmetric $SU(N)$ Yang-Mills theory at finite temperature and compactified on a circle
for a range of $N=2,3,4$. In particular we have focused on the spatial Polyakov
line as an order parameter for a large $N$ deconfining phase transition. 
Our simulations are consistent with the existence of a single transition curve in the 2d parameter
space spanned by the two dimensionless couplings, $r_x, r_\tau$ which give the size of the thermal and spatial circle in units of the YM coupling. 

At high temperature, $r_\tau^3 \ll r_x$, the simulations are consistent with the previously predicted behavior that the transition curve goes as $r_x^3 = 1.35 r_\tau$. At strong coupling, $1 \ll r_\tau$, the transition is conjectured to be the holographic dual of a first order Gregory-Laflamme
phase transition, with the transition curve going a $r_x^2 = c_{crit} r_\tau$, with $c_{crit}$ an order one constant obeying the constraint $c_{crit} > 2.29$. Our simulations are consistent with this parametric behavior, and we have used the $N=3,4$ data to estimate the position of the large $N$ transition, determining 
$c_{crit}$ to be very close to the GL instability value $\simeq 2.29$. 
This gives the ratio of the Gregory-Laflamme phase transition and dynamical instability temperatures
$T_{GL phase}/T_{GL instability}$ to be  close to one. 
Since the dual localized black hole solutions have not been constructed, this constitutes a prediction for these non-trivial gravity solutions, which hopefully will be tested by their construction in the near future.
 
%
\acknowledgments
We thank Paul Romatschke for bringing to our attention the error in normalization of the lattice coupling in previous versions of this work.
SC and AJ are supported in part by the US Department of Energy under grant DE-FG02-85ER40237. TW is supported by a STFC advanced fellowship and a Halliday award. Simulations were performed using USQCD resources at Fermilab.
%

%
\bibliographystyle{JHEP}
\bibliography{ref}
%

\end{document}